\documentclass{vgtc}                          % final (conference style)
\graphicspath{{figures/}{pictures/}{images/}{./}} % where to search for the images

\usepackage{times}                     % we use Times as the main font
\usepackage{tabu}                      % only used for the table example
\usepackage{lipsum}                    % used to generate placeholder text
\usepackage{mwe}                       % used to generate placeholder figures
\usepackage{graphicx}
\usepackage{tabularx}
\usepackage{array}
\usepackage{longtable}
\usepackage{wrapfig}
\usepackage{caption}
\newcolumntype{C}{>{\centering\arraybackslash}X}
\usepackage{booktabs, multirow} % for borders and merged ranges
\usepackage{xcolor,colortbl} % for cell colors
\usepackage{changepage,threeparttable} % for wide tables

\usepackage{xcolor}

\usepackage{float}

\usepackage{mathptmx}                  % use matching math font

\usepackage{amsmath}
\usepackage{enumitem}
\usepackage[most]{tcolorbox}

\definecolor{delcolor}{HTML}{8B0000}      % dark red
\definecolor{addcolor}{HTML}{00008B}      % dark blue
\definecolor{revisecolor}{HTML}{CC6600}   % dark orange

\onlineid{0}

\vgtccategory{Research}

\vgtcinsertpkg

\title{How Does Title Framing Influence Pattern Identification in Line Charts?}

\author{Jasmine Lim\thanks{e-mail: \{jasmine.t.lim, pandey, quadri\}@ou.edu} \\ %
        \scriptsize University of Oklahoma %
\and Tapendra Pandey\footnotemark[1] \\ %
     \scriptsize University of Oklahoma %
\and Arran Zeyu Wang\thanks{e-mail: zeyuwang@cs.unc.edu} \\ %
     \scriptsize UNC-Chapel Hill %
\and Sungahn Ko\thanks{e-mail: sungahn@postech.ac.kr} \\ %
     \scriptsize POSTECH %     
\and Ghulam Jilani Quadri\footnotemark[1] \\ %
     \scriptsize University of Oklahoma}

\teaser{
  \centering
  \includegraphics[width=\linewidth]{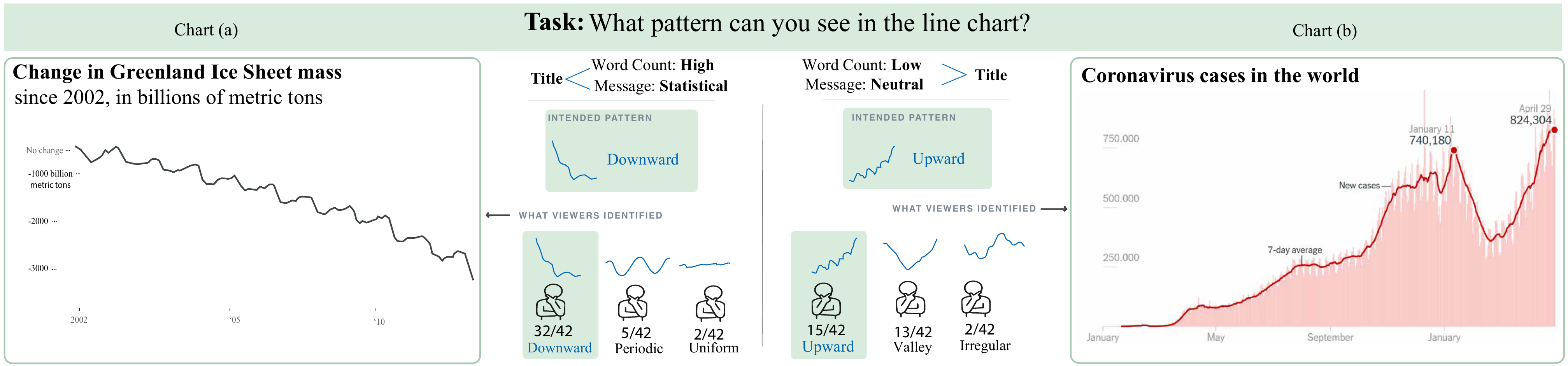}
  \caption{We conducted a user study eliciting quick pattern identification from line charts. Participants reported different patterns for the same chart: for Chart (a), 
  32 of 42 participants reported the intended pattern (\textit{downward}), while 
  15 of 42 participants reported the intended pattern (\textit{upward}) for
  Chart (b). We investigate how title characteristics, word count (low, medium, high), and intended message (neutral, affective, statistical) influence which pattern viewers identify.}
  \label{fig:teaser}
}

\abstract{
    Visual data communication in digital media is increasingly characterized by short attention spans and snapshot-based viewing, often employing line charts to convey trends and patterns.
   ~Among all visual elements, titles are 
   crucial elements that can shape how viewers interpret visual information and form chart takeaways.
   In this study, we examine how title characteristics, particularly title word count and intended message, influence people's pattern identification in single-class line charts.
   Participants viewed 50 line charts collected from online news media and identified the pattern they perceived.
   Our results demonstrate that both title word count and intended message significantly influence viewers' pattern identification.
   Our findings highlight the importance of title framing in shaping quick-view pattern takeaways and supporting effective visualization communication.

} % end of abstract


\begin{document}


% Reviewer 1
% - Clarify the exact instructions and pattern definitions shown to participants before the study.
% - Define what counted as a chart title, including how subtitles and nearby in-chart text were treated.
% - Explain why stimuli 16 and 43 were considered to have titles despite their unconventional placement.
% - Justify combining titles and subtitles in the word-count measure and acknowledge that subtitles may contribute different information.
% - Soften the interpretation of high word count as “richer framing”; distinguish title length from contextual and interpretive information.

% Reviewer 2
% - Add a table showing the stimulus distribution across word count, intended message, and intended pattern.
% - Acknowledge that title characteristics and visual structure may covary, creating possible pattern-difficulty effects.
% - Scale the statistical claims by noting that responses are nested within participants and charts.
% - Report an effect-size measure alongside the chi-squared results.
% - Examine time-to-response by title word count, if available, to assess whether longer titles affect how participants allocate the 15-second viewing period.
% - Discuss alternative explanations for the word-count effect, especially richer semantic information versus reduced chart-inspection time.
% - Fix the paper title grammar/question form. 

\firstsection{Introduction}
\maketitle
% \fix{revision summary in main.tex}
% \fix{reviewer 1 asks for stimulus distribution table}
% \fix{determine analyzing response time}
% \flushbottom
\enlargethispage{0.5in}
Line charts, which date back to William Playfair~\cite{playfair1801commercial}, are commonly used for visualizing time-series and continuous data in online news media to communicate trends, comparisons, and changes over time \cite{quadri2021survey,franconeri21}.
% In modern digital spaces, visual data communication is increasingly characterized by brief attention spans and snapshot-based skimming, where readers frequently encounter visualizations embedded in fast-paced news articles. 
Within digital news spaces, readers spend a few seconds viewing a visualization before proceeding, forming quick takeaways rather than carefully analyzing the underlying data \cite{kim2024ThumbnailArticles}. Capturing these quick takeaways represents a form of intuitive insight building beyond traditional graphical perception tasks, called high-level visualization comprehension \cite{quadri2024do, jeon2026llm}.
% Under these conditions, viewers typically form quick judgments on line charts by identifying high-level, salient visual structures, such as sudden trends, sharp peaks, and steep declines, to anchor their understanding of the underlying data. 
% Under these conditions, 
% Viewers rely on salient visual structures, such as sudden increases, sharp peaks, and steep declines, when forming quick takeaways from a chart in the wild.
% 
% Pattern identification represents one of the primary ways readers form quick takeaways from line charts, making it a fundamental component of visualization communication.
Viewers often rely on salient visual structures, such as sudden increases, sharp peaks, and steep declines, to identify patterns and form quick takeaways from line charts, making pattern identification a fundamental component of visualization communication.

% However, charts in online media rarely appear in isolation.
% Alongside visual structure, titles provide readers with contextual information that can frame how a chart is understood \cite{kim2021understandingReadersIntegrateChartsCaptions}.
% However, readers rarely rely on visual structure alone. Charts in online media are typically accompanied by textual information that can frame how a chart is interpreted.
%
However, readers rarely rely on visual structure alone, as charts in online media are typically accompanied by textual information that can frame their interpretation.
%lundgard2021accessibleVisualizationNaturalLanguage 
% Readers often form quick judgments on line charts by identifying high-level patterns such as trends, peaks, and declines
% . These identified patterns 
% guiding viewers to interpret the visualized data. 
% As the primary textual feature accompanying many news charts, t
Titles are often the primary textual feature in these charts and can influence how viewers interpret visualized information~\cite{stokes2022striking}. Therefore, understanding the role of title framing in pattern identification is significant for effective visual communication.\\
\indent Titles vary in both the messages they communicate and the amount of text they contain to communicate those messages, representing two common types of textual framing observed in real-world news visualizations.
Textual cues 
% have been shown to 
influence chart comprehension, recall, judgment, and takeaways \cite{kong2018slantsTitles,stokes2022striking, kim2021chartsCaptions}. 
% In addition, textual framing can communicate different messages, ranging from descriptions of visible patterns to broader contextual explanations and insights \cite{stokes2022striking}. However, it remains unclear how title characteristics influence viewers' pattern identification during quick viewing of real-world news charts.
Although previous studies have examined individual aspects of textual framing using controlled chart stimuli, it remains unclear how title characteristics influence viewers' pattern identification during quick viewing of real-world news charts.
In this work, we investigate how two characteristics of title framing, intended message and title word count, influence pattern identification in single-class line charts collected from online news media. Our stimuli contain titles that communicate neutral, statistical, and affective messages~\cite{kong2018slantsTitles}.
We conducted a user study with 47 participants, who viewed 50 charts and identified the pattern from seven predefined categories (see Section \ref{sec-experiment}) 
% We aim 
to investigate how title framing in single-class line 
% charts affects readers' 
charts influences pattern takeaways.\\
\indent We found that title word count and intended message significantly influence pattern identification. 
% However, visually salient structures helps the viewer identify the pattern, even when the title framing remains neutral and short. 
Visually salient structures often guided participants toward the intended pattern even when title framing was neutral or brief.
We further observed 
% substantial 
variability across charts, suggesting that textual framing and salient visual structure jointly contribute to the pattern 
% interpretation. 
identification.
% , although visually salient structures often remain dominant.
%
%
% Our findings provide empirical evidence on the role of title framing in shaping chart takeaways and offer guidance for designing public-facing visualizations that align textual cues with designers' communicative objectives. This understanding is especially important in quick-view contexts, where readers often glance at charts embedded in news articles or social media and form quick takeaways without closely inspecting the underlying data. By demonstrating how title framing affects pattern identification, this work provides evidence-based design guidelines to help visualization designers maximize clarity and minimize unintended interpretations. Ultimately, this serves as a foundational investigation of textual framing in visualization, expanding our understanding of how textual elements and visual structure collectively influence chart interpretation.
% Our findings provide empirical evidence that title framing contributes to viewers' quick-view pattern identification while highlighting the complementary role of salient visual structure. This understanding is especially important in quick-view contexts, where readers often glance at charts embedded in news articles without closely inspecting the underlying data. By demonstrating how title framing affects pattern identification, this work provides an evidence-based foundation for designing visualizations that maximize clarity and reduce unintended interpretations of pattern. 
%
Our findings provide empirical evidence that title framing influences viewers' pattern identification in line charts while highlighting the complementary role of salient visual structures. These findings are particularly relevant in quick-view settings, where readers 
% often 
encounter charts embedded in online media and form impressions without closely examining the underlying data. By characterizing the influence of title framing on pattern identification, our work contributes to a better understanding of how visualizations can support clearer 
% communication and mitigate divergent interpretations. 
communication and more consistent interpretation.
% More broadly, 
Overall, this work serves as an initial investigation of textual framing in chart titles to understand how textual elements and visual structure jointly influence chart interpretation.

\label{sec-intro}

% \vspace{-8pt}

\section{Related Works}

% \begin{figure}[b]
%   \centering
%   \includegraphics[width=\columnwidth]{figures/trend_.pdf}
%   \vspace{-15pt}
%   \caption{Shows the example stimuli used in our study, with the title word count -- \textbf{low}, intended message -- \textbf{affective}, and intended pattern -- \textbf{upward}.}
%   \vspace{-15pt}
%   \label{fig:trend}
% \end{figure}

%Related work spans three areas relevant to our study: visualization communication through line charts, visual pattern perception, and textual framing. Prior research has examined how line charts communicate information, how viewers identify patterns from visual structure, and how textual elements shape chart interpretation. 

We situate our work at the intersection of three research areas: visualization communication, visual pattern perception, and textual framing. Previous studies have examined how line charts communicate information, how viewers identify patterns from salient visual structures, and how textual elements shape interpretation.
% Building on these research areas, we investigate how title framing and salient visual structure jointly influence quick-view pattern identification in real-world single-class line charts.

\subsection{Line Charts in Visualization Communication}

% Charts have served as a medium for communicating quantitative information since the introduction of statistical graphics \cite{playfair1801commercial}.
% 
Line charts are widely used to visualize time-series data, enabling viewers to observe key patterns \cite{mussweiler2003goes, assfalg2009periodic}.
% Visualization research increasingly views charts as communication artifacts designed to help viewers construct meaningful takeaways from data rather than simply perform isolated perceptual tasks \cite{franconeri21,quadri2024do}. Accordingly, 
Recent work has emphasized evaluating visualizations based on viewers' high-level comprehension and semantic understanding, examining how the intended message conveyed by visualizations aligns with the interpretations formed by their audience \cite{lundgard2021Semantic,quadri2024do, jeon2026llm}. 
This perspective is particularly relevant for public-facing visualizations, where readers often encounter charts with limited attention and rely on immediate interpretations \cite{kim2024ThumbnailArticles}. Understanding the factors that shape these interpretations is therefore central to effective visualization communication.

\vspace{-5pt}
\subsection{Visual Pattern Perception in Line Charts}
% how do people identify patterns from a visual structure
% change first sentence
% Rather than extracting individual data values, viewers often summarize line charts by identifying higher-level visual patterns that capture the overall behavior of a time series. 
% Prior work has shown that line chart interpretation relies on judgments of aggregate properties such as trends, averages, and similarity, rather than precise point-by-point comparisons \cite{correll2012AveragesTimeSeries,albers2014taskAggTimeSeries,gogolou2019comparingSimilarityPerception, quadri2021survey}. 
% Viewers typically interpret line charts by summarizing aggregate properties such as trends, averages, and similarity, rather than comparing individual data points \cite{correll2012AveragesTimeSeries,albers2014taskAggTimeSeries,gogolou2019comparingSimilarityPerception,quadri2021survey}.
% Viewers generally interpret line charts by summarizing aggregate properties such as trends and averages rather than comparing individual data points \cite{correll2012AveragesTimeSeries,albers2014taskAggTimeSeries}. 
An extensive body of work on perceptual and cognitive processes has informed our understanding of how people estimate values in visualizations, beginning with seminal work by Cleveland and McGill~\cite{cleveland1984graphical}. More recent research has expanded beyond the extraction and comparison of individual values to examine graphical perception in more holistic statistical tasks~\cite{franconeri21,quadri2021survey, schloss2025perceptual}.
Viewers generally make aggregate judgments about trends and averages instead of relying on individual data values \cite{correll2012AveragesTimeSeries,albers2014taskAggTimeSeries}.
Similarly, studies indicate that human perception of temporal trends, continuous changes, and comparisons
% between time series 
is driven by the overall similarity of visual patterns rather than exact point-by-point correspondence, a pattern clearly visualized by line charts \cite{gogolou2019comparingSimilarityPerception,wang2018linegraphVisualizingTrends,jeon2026llm}. 
More recent studies further demonstrate that viewers prioritize visually salient structures when interpreting line charts, consistently emphasizing features such as overall trends, peaks, and valleys while abstracting away less salient variations and noise \cite{proma2025evaluatinglinechartstrategies,proma2025stenography, rosen2021smoothing}. At the same time, other work shows that visual salience, graphical conventions, and perceptual biases can systematically influence how viewers interpret line charts, directing attention toward particular features and affecting the conclusions they derive from the same underlying data \cite{matzen2018saliencymodel,woodin2022graphicalconvention,moritz2024averageEstimatesBiased}. Overall, these studies demonstrate that a chart's visual characteristics strongly shape pattern identification. 

Although viewers often agree on broad trends, line charts frequently contain multiple visually salient features that support different interpretations. The features that attract attention can shape which aspects of a chart are emphasized, allowing different viewers to derive distinct interpretations from the same underlying data \cite{oral2024decouplingJudgement,xiong2020biasedPositionLine}. These findings suggest that pattern identification emerges from an interaction between the underlying data and the visual structures that guide attention, creating opportunities for contextual information, such as titles and captions, to influence interpretation~\cite{stokes2022striking, kim2021chartsCaptions}. However, despite growing evidence that contextual information can influence interpretation, comparatively little attention has been paid to how textual elements interact with visual structure to shape viewers' perceived patterns.

\subsection{Textual Framing in Visualization}

Text accompanying a chart is an integral component of communication, providing semantic context that complements graphical representations and guides viewers' interpretation of data \cite{lundgard2021Semantic,stokes2022striking,stokes2025analysistextfunctionsinformation}. These textual elements, such as titles, captions, and annotations, work together with visual encodings to communicate intended messages and shape the takeaways viewers derive from visualizations \cite{kim2021chartsCaptions,zhu2022captionsaffectvisualizationreading,stokes2024roleOfAnnotations}. Among these textual elements, titles establish readers' expectations and frame their interpretation before they inspect the visualization, and different framing strategies can influence the messages viewers remember and the conclusions they draw from charts \cite{wanzer2021titlerole,kong2018slantsTitles,kong2019recallMisalignment,lauer2023DeceptiveTactics}. 
% While prior findings explain textual framings' influence on visualization interpretation, they have primarily examined textual elements independently or focused on overall chart comprehension. 
Despite these findings, prior work has primarily examined textual elements independently or focused on overall chart comprehension.
It remains unclear how naturally occurring title framing and salient visual structure together influence the visual patterns viewers perceive in real-world line charts during brief viewing.

\label{sec-related}

% \vspace{-8pt}

\section{Study Design}
\begin{figure}[t]
  \centering
  \includegraphics[width=\columnwidth]{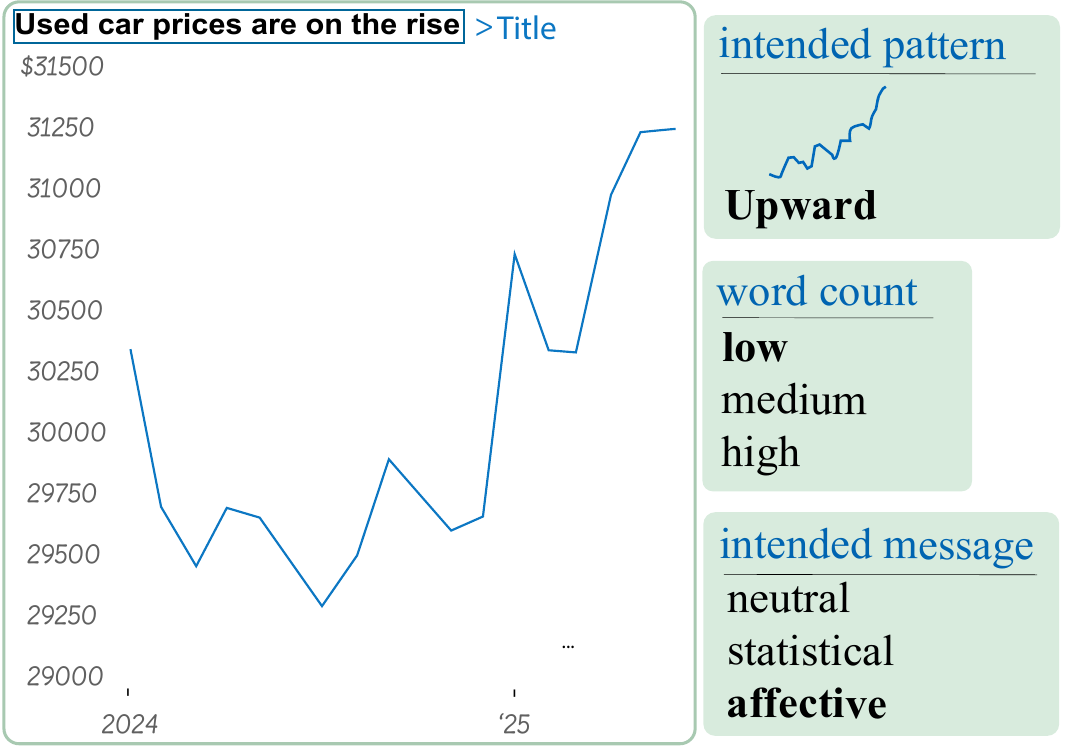}
  \vspace{-15pt}
  \caption{The example stimuli used in our study;
  % with the
  % title "\textbf{Used car prices are on the rise}," 
  Title word count -- \textbf{low}, intended message -- \textbf{affective}, intended pattern -- \textbf{upward}.}
  \vspace{-19pt}
  \label{fig:trend}
\end{figure}

To examine how title framing influences quick-view pattern identification in line charts,
% we conducted a web-based user study using real-world charts collected from online news media. 
we conducted a user study to elicit the visual pattern that viewers identify after viewing each line chart. We first define the line-chart patterns, collect the stimulus set, characterize the titles of the stimulus charts, and finally describe the study procedure.\\

\vspace{-5pt}
\noindent \textbf{Patterns:}
% \label{sec-pattern}
To capture common patterns observed in real-world news visualizations and provide a consistent 
vocabulary for pattern identification, we defined seven possible intended patterns (\textit{upward, downward, peak, valley, periodic, uniform}, and \textit{irregular}) \cite{proma2025evaluatinglinechartstrategies}.

\label{sec-pattern}
\begin{figure}[H]
\raggedright
\footnotesize
\vspace{-8pt}

\noindent
\begin{minipage}[t]{0.16\columnwidth}
    \raggedright
    \vspace{0pt}
    \includegraphics[width=0.7\linewidth]{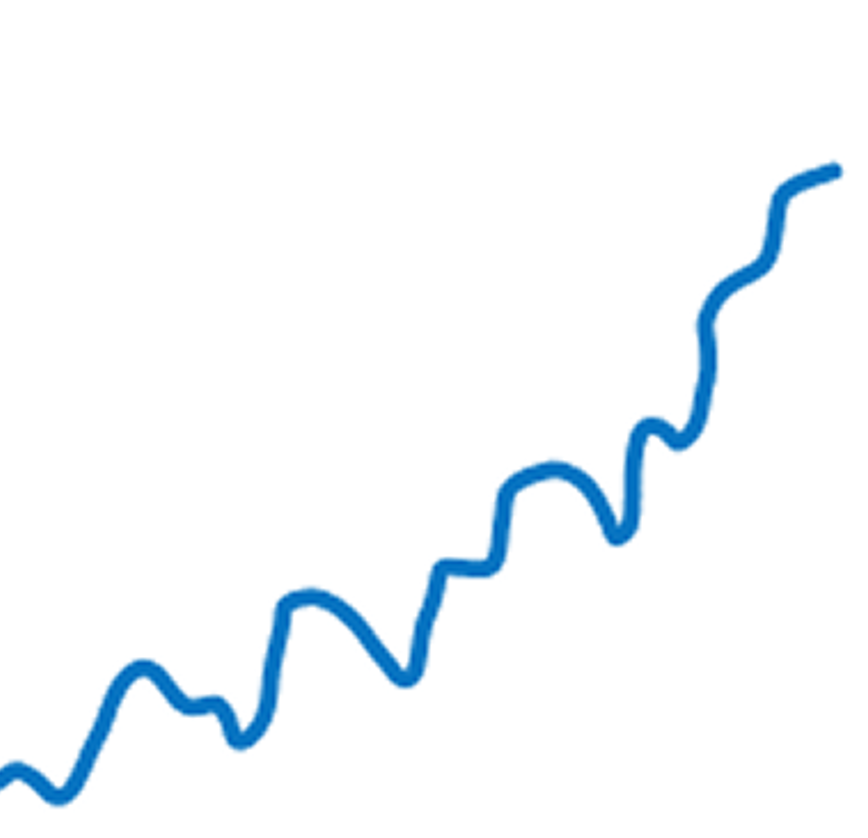}
\end{minipage}%
\hspace{-5pt}%
\begin{minipage}[t]{0.80\columnwidth}
    \vspace{2pt}
    % \textit{Increasing} -- Increasing trend over time.
    \textit{Upward} -- The line shows an increasing trend over time. Although small fluctuations may occur, later values are generally higher than earlier values.
    
\end{minipage}

\vspace{3pt}

\noindent
\begin{minipage}[t]{0.16\columnwidth}
    \raggedright
    \vspace{2pt}
    \includegraphics[width=0.7\linewidth]{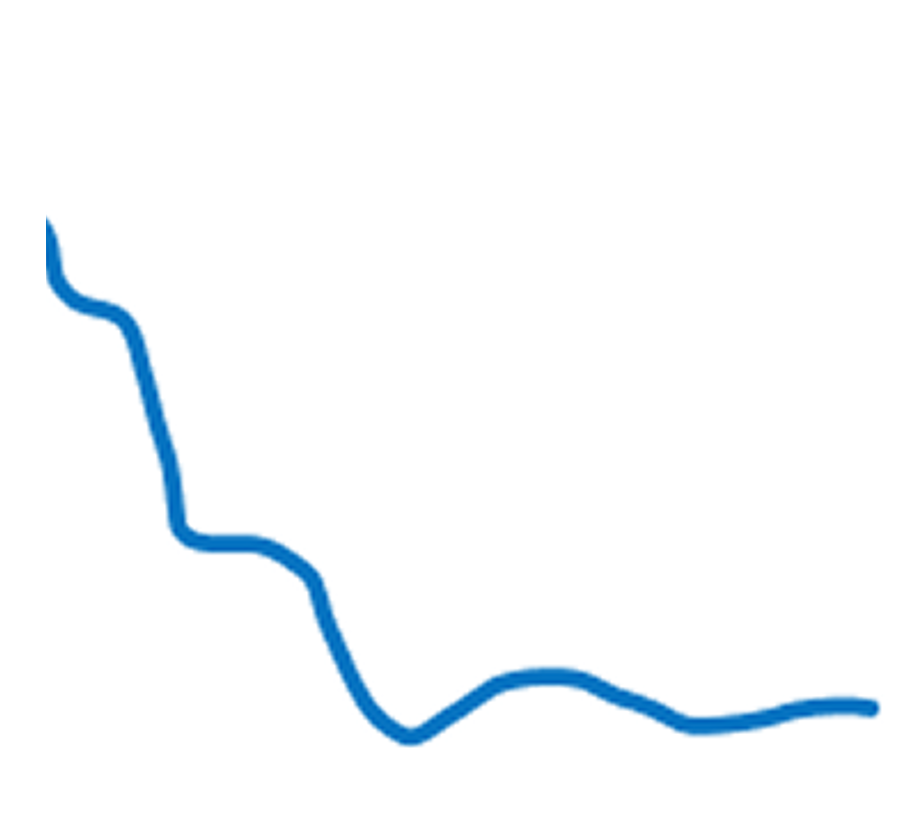}
\end{minipage}%
\hspace{-5pt}%
\begin{minipage}[t]{0.80\columnwidth}
    \vspace{2pt}
    % \textit{Decreasing} -- Decreasing trend over time.
    \textit{Downward} -- The line shows a decreasing trend over time. Despite possible short-term fluctuations, later values are generally lower than earlier values.
    
\end{minipage}

\vspace{-3pt}

\noindent
\begin{minipage}[t]{0.16\columnwidth}
    \raggedright
    \vspace{-5pt}
    \includegraphics[width=0.7\linewidth]{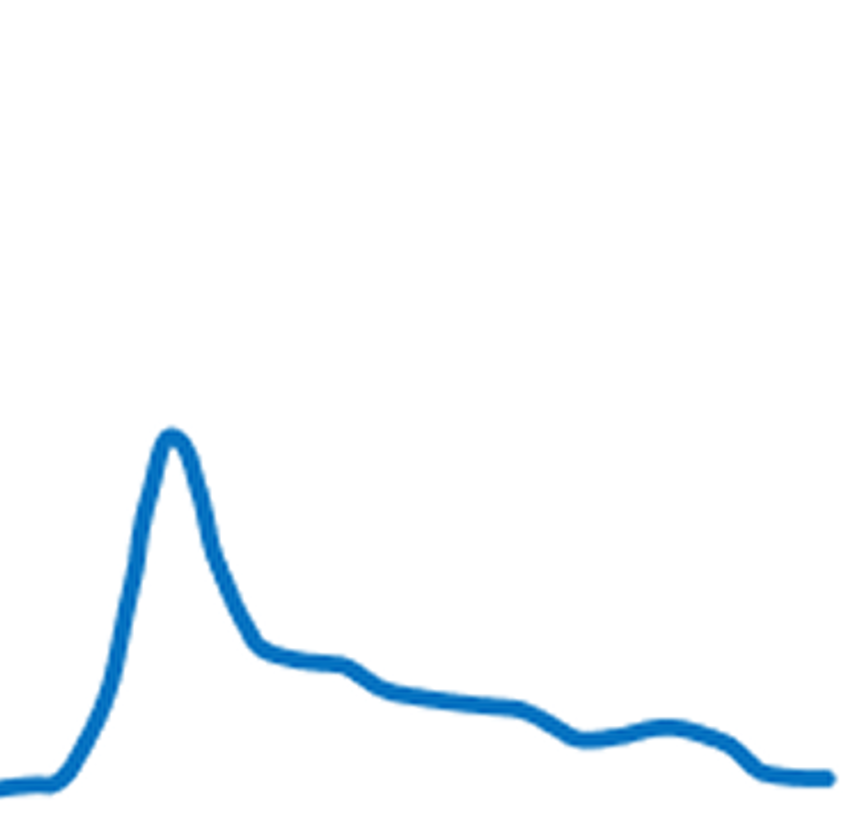}
\end{minipage}%
\hspace{-5pt}%
\begin{minipage}[t]{0.80\columnwidth}
    \vspace{2pt}
    % \textit{Peak} -- Prominent rise leading to a local maximum.
    \textit{Peak} -- The line rises toward a prominent local maximum. The peak represents a visually salient high point within the displayed time period.
    
\end{minipage}

\vspace{3pt}

\noindent
\begin{minipage}[t]{0.16\columnwidth}
    \raggedright
    \vspace{0pt}
    \includegraphics[width=0.7\linewidth]{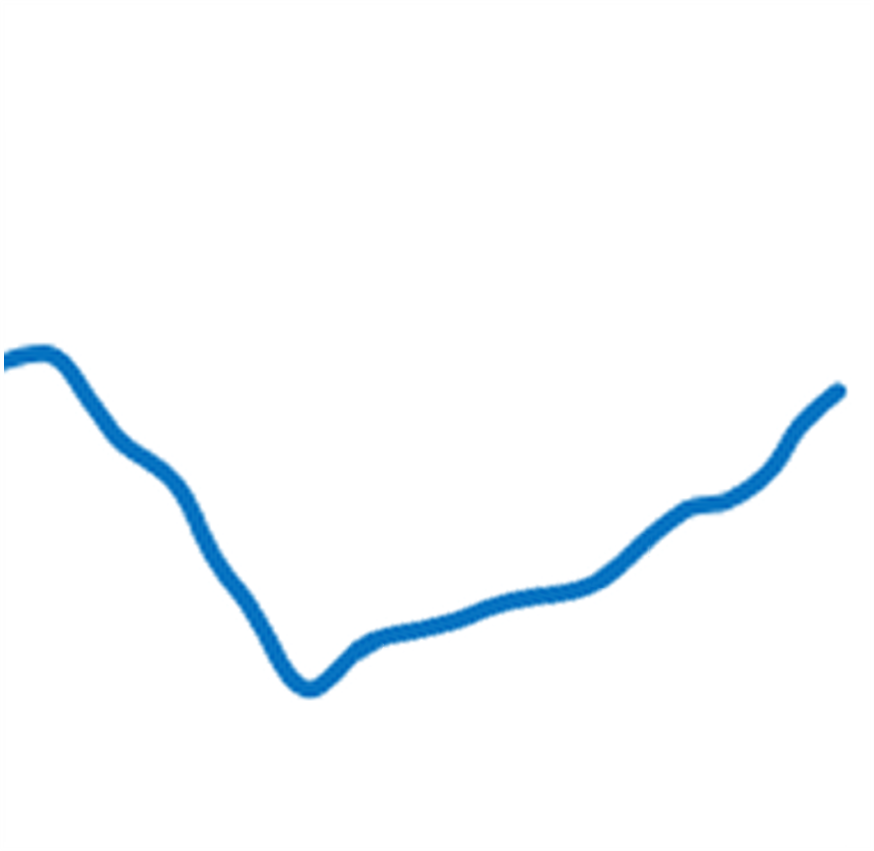}
\end{minipage}%
\hspace{-5pt}%
\begin{minipage}[t]{0.80\columnwidth}
    \vspace{2pt}
    % \textit{Valley} -- Prominent fall leading to a local minimum.
    \textit{Valley} -- The line falls toward a prominent local minimum. The valley represents a visually salient low point within the displayed time period.
    
\end{minipage}

\vspace{-3pt}

\noindent
\begin{minipage}[t]{0.16\columnwidth}
    \raggedright
    \vspace{0pt}
    \includegraphics[width=0.7\linewidth]{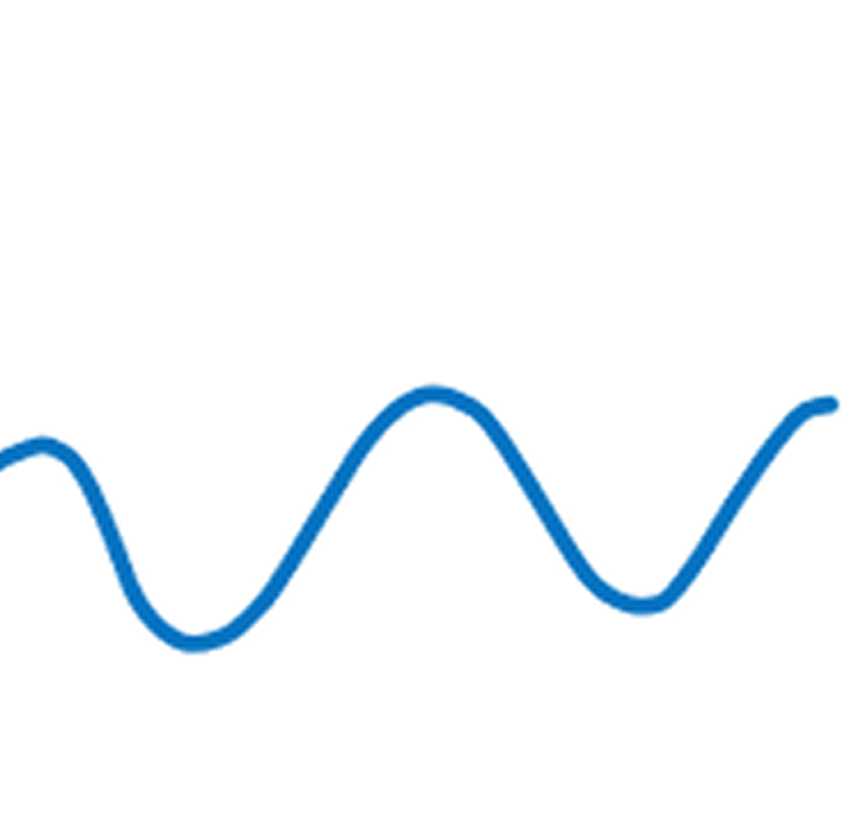}
\end{minipage}%
\hspace{-5pt}%
\begin{minipage}[t]{0.80\columnwidth}
    \vspace{5pt}
    % \textit{Periodic} -- Regular rise and fall over time.
    \textit{Periodic} -- The line shows repeated rises and falls over time. These fluctuations occur in a regular or recurring pattern.
    
\end{minipage}

\vspace{-3pt}

\noindent
\begin{minipage}[t]{0.16\columnwidth}
    \raggedright
    \vspace{0pt}
    \includegraphics[width=0.7\linewidth]{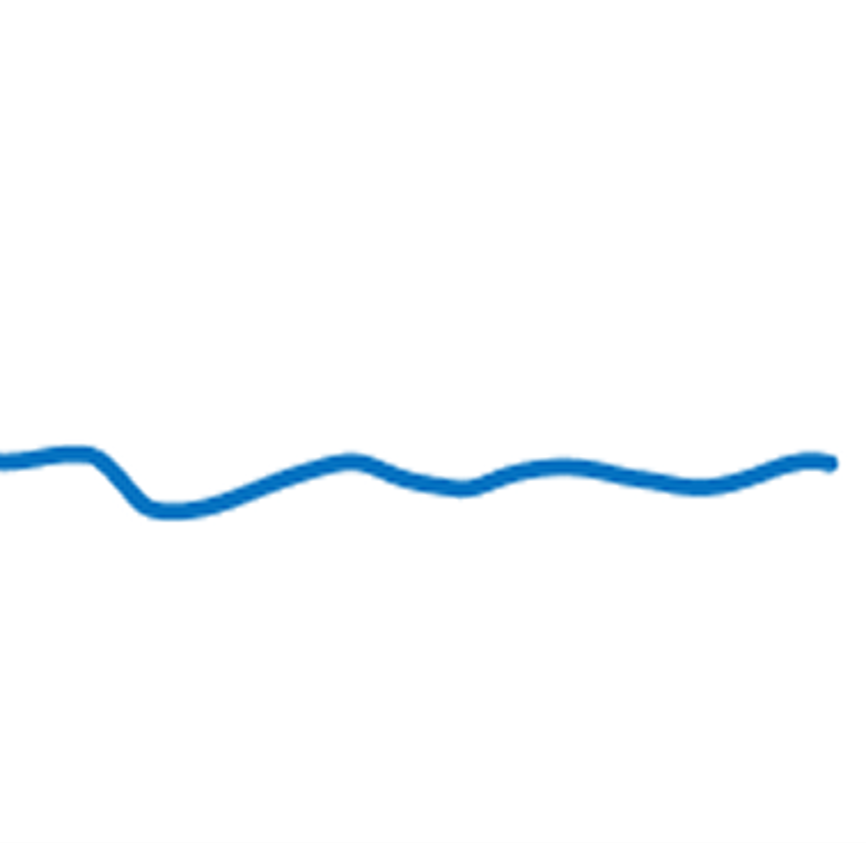}
\end{minipage}%
\hspace{-5pt}%
\begin{minipage}[t]{0.80\columnwidth}
    \vspace{5pt}
    % \textit{Uniform} -- Constant over time with little to no variation.
    \textit{Uniform} -- The line remains relatively constant throughout the displayed period. Values show little to no meaningful variation over time.
\end{minipage}

\vspace{0pt}

\noindent
\begin{minipage}[t]{0.16\columnwidth}
    \raggedright
    \vspace{0pt}
    \includegraphics[width=0.7\linewidth]{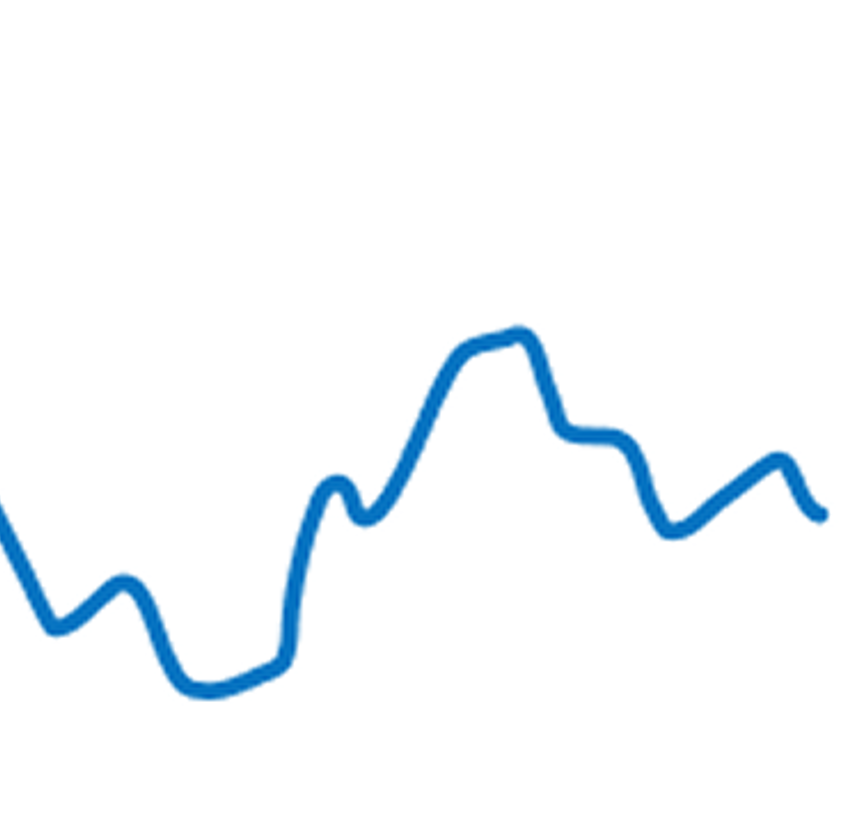}
\end{minipage}%
\hspace{-5pt}%
\begin{minipage}[t]{0.80\columnwidth}
    \vspace{5pt}
    % \textit{Irregular} -- No dominant trend present.
    \textit{Irregular} -- The line fluctuates without a clear dominant direction or recurring structure. No single upward, downward, peak, valley, periodic, or uniform pattern adequately characterizes the chart.

\end{minipage}

% \caption{The seven intended patterns identified in single-class line charts. Each pattern represents a distinct trend used to characterize viewers' pattern identification throughout the study.}
\label{fig:patterns}
\vspace{-5pt}

\end{figure}

\noindent \textbf{Stimuli:} We collected 50 single-class static line charts 
from \href{https://nytimes.com}{The New York Times} ($n=24$) and \href{https://wsj.com}{Wall Street Journal} ($n=26$)
% that have a visible title and a clearly identifiable statistical pattern.
that contained identifiable title text and a clearly identifiable pattern.
\autoref{tab:50} summarizes the distribution of stimuli across title word count, intended message, and intended pattern.
% \add{Across our stimuli, the intended patterns are: upward $(n=12)$, downward $(n=3)$, peak $(n=12)$, valley $(n=9)$, uniform $(n=0)$, irregular $(n=12)$}.

\noindent \textbf{Title Characteristics:}
\noindent 
% Titles are ... position in chart.. all text spatially close to header is title....
% define title characteristics first, then title definition.
% for figure 2, show that it has the title in section 3
% then describe intended message, then word count, then in this figure 2 - it has word count, etc
%
Titles are design elements in charts that frequently appear at the top of the visualization. They
% The title accompanying a chart 
provide contextual information that frames how readers interpret the underlying data.~Chart titles therefore act as primary text that introduces or frames the intended message.
We characterize title framing through two features:
% We characterize the textual framing of each chart through two title characteristics:
%
a) title word count and b) intended message, as illustrated in \autoref{fig:trend}.
% \add{where the title is "Used car prices are on the rise."}. 
Title word count refers to the number of words in the title, including prepositions, but excluding symbols. 
% \add{Subtitles and} other
Textual elements, such as axis labels, source information, and annotations within the chart, were not included in the word count.
We categorized the title word count as low (1-5 words), medium (6-10 words), and high (11+ words). The intended message 
% represents the designer’s intended framing of how the data and pattern should be interpreted 
characterizes how the chart's title frames the underlying data and its interpretation
\cite{bai2026feeldesignersbringemotions}. We categorized the intended message as \textit{neutral}, \textit{statistical}, or \textit{affective} \cite{kong2018slantsTitles}. \textit{Neutral} titles describe the topic with minimal interpretation (see \autoref{fig:teaser} Chart (b)). \textit{Statistical} titles emphasize numeric values, magnitudes, or quantitative comparisons (see \autoref{fig:teaser} Chart (a)). \textit{Affective} titles contain evaluative or emotionally suggestive language without any statistics (see \autoref{fig:trend}).
We identified the intended pattern for each line graph by 
% analyzing
considering its visual structure 
% alongside
together with the corresponding article's title, caption, and content.
% , treating
% The pattern most strongly 
% % conveyed
% supported by both the visualization and the narrative was treated as the ground truth for evaluating participants' identified patterns.
% % 
% The intended pattern (see~\autoref{fig:trend}) served as the ground truth for evaluating the correctness of participants' identified patterns.
The pattern most strongly supported by both the visualization and narrative served as the ground truth for evaluating participants’ responses (see~\autoref{fig:trend}).

% change to active voice
% improve flow?
\noindent \textbf{Procedure:} We conducted a priori power analysis using G*Power 3.1 for a within-subjects repeated measures ANOVA with ($f = 0.25$), ($\alpha$ = .05), ($1-\beta = .80$), and ($r = .50$) among repeated measures indicating a required sample size of $n = 28$. To ensure sufficient statistical power, we recruited 47 participants, including 22 via university mailing lists and online announcements and 25 via Prolific.
Participants were compensated at a rate of \$12 per hour.
We conducted a web-based study where participants viewed one chart at a time and identified the pattern they perceived. Participants reported their age and educational background.
Before beginning the study, participants reviewed the definitions of all seven predefined patterns.
We presented the descriptions once before the first experimental trial only.
Each participant completed 50 trials. In each trial, a chart was displayed for up to 15 seconds while participants answered the question: “\textit{What pattern can you see in the line chart}?”
% Participants could respond at any time during this interval and complete the specified chart trial. 
Participants selected one pattern label from all seven predefined pattern categories, which were presented as radio-button options in a fixed order.
% multiple-choice options using radio buttons.
We included two attention check trials throughout the study to eliminate random clicks.

\label{sec-experiment}

% \vspace{-10pt}

\section{Analysis and Results}
We conducted quantitative and qualitative analyses of title word count and intended message to examine whether title framing influences pattern identification.
We excluded responses from five participants who failed at least one attention check ($n=2$) or reported color blindness ($n=3$), resulting in a final sample of 42 participants.
Across these 42 participants, we collected 2100 responses. We excluded an additional 13 responses due to timeouts, resulting in 2087 valid responses for analysis. 
Two independent coders assigned an intended pattern label to each chart from the predefined pattern categories. Inter-rater reliability was assessed using Cohen’s Kappa, yielding substantial agreement ($\kappa = 0.85$). 
The same coders qualitatively reviewed all 50 charts to identify visually salient structures. We resolved disagreements through discussion to produce a final consensus label.
% and resolved disagreements through discussion before reaching consensus.

% \renewcommand{\arraystretch}{0.9}
% \begin{table}[!htp]\centering
% \caption{Log-linear model results for title word count and intended message. 
% Significant effects are indicated by \textbf{bold} text. \add{This model treats each response as
% independent, although responses are nested within 42 participants and 50 charts.}
% }
% \vspace{-10pt}
% \begin{tabular}{lrrr}
% \toprule
% \textbf{Comparison} & \textbf{$\Delta \chi^2$} & \textbf{df} & \textbf{p-value} \\
% \midrule
% Overall model improvement & 206.04 & 28 & \textbf{$<$ .001} \\
% Title Word Count & 129.09 & 14 & \textbf{$<$ .001} \\
% Intended Message & 76.94 & 14 & \textbf{$<$ .001} \\
% \bottomrule
% \end{tabular}
% \label{tab:loglinear}
% \vspace{-5pt}
% \end{table}

\renewcommand{\arraystretch}{0.9}
\begin{table}[!t]\centering
\caption{Distribution of the 50 chart stimuli across title word count, intended message, and intended pattern categories.}
{\footnotesize
\setlength{\tabcolsep}{3.5pt}
\renewcommand{\arraystretch}{1.2}

\begin{tabular}{
p{1.0cm}
p{0.4cm} |
p{1.2cm}
p{0.4cm} |
p{3.8cm}
}
\hline

\multicolumn{1}{>{\raggedright\arraybackslash}m{1.0cm}}{\textbf{Word Count}} &
\multicolumn{1}{>{\raggedright\arraybackslash}m{0.4cm}|}{\textbf{N}} &
\multicolumn{1}{>{\raggedright\arraybackslash}m{1.2cm}}{\textbf{Intended Message}} &
\multicolumn{1}{>{\raggedright\arraybackslash}m{0.4cm}|}{\textbf{N}} &
\multicolumn{1}{>{\raggedright\arraybackslash}m{3.8cm}}{\textbf{Intended Pattern (N)}} \\
\hline

Low    & 18 & Neutral     & 19 & Upward (12), Downward (3) \\
Medium & 22 & Statistical & 23 & Peak (12), Valley (9), Periodic (2) \\
High   & 10  & Affective   & 8  & Uniform (0), Irregular (12) \\

\hline
\end{tabular}}

\label{tab:50}
% \vspace{-15pt}
\end{table}

Quantitative analysis measures the overall effects of title word count and intended message across the entire dataset, while qualitative analysis examines chart-by-chart variations to explore how visual structure shapes those effects. We use seven pattern categories to ensure that our findings capture diverse real-world news visualizations and compare participants' identified patterns with intended patterns. Because different line charts vary in visual salience, tracking performance across these categories helps reveal when textual framing acts as an interpretive cue and when the underlying visual structure alone guides the viewer's pattern takeaways.

\vspace{-5pt}
\subsection{Quantitative Analysis} 
To examine whether textual characteristics affected pattern identification, we fitted a hierarchical log-linear model. Log-linear analysis examines relationships among categorical variables in multi-way contingency tables and tests whether the distribution of identified pattern labels
varies across levels of the textual characteristics \cite{agresti2013cda}.
We compared an independence model assuming that pattern identification, title word count, and intended message were mutually independent ($\chi^2 = 11626.45$) with a model including two interactions: pattern identification $\times$ title word count and pattern identification $\times$ intended message. Including both interactions significantly improved model fit ($\Delta\chi^2 = 268.48$, $df = 28$, $p < .001$), indicating that the distribution of identified pattern labels differed across title word count and intended message.
% , shown in \autoref{tab:loglinear}.

Both the pattern identification $\times$ title word count interaction ($\Delta\chi^2 = 155.06$, $df = 14$, $p < .001$, Cramér's $V = .19$) and the pattern identification $\times$ intended message interaction ($\Delta\chi^2 = 113.42$, $df = 14$, $p < .001$, Cramér's $V = .16$) significantly improved model fit, 
% demonstrating that both title word count and intended message independently influence quick-view pattern identification 
indicating that identified pattern distributions varied across both title word count and intended message categories.
% in single-class line charts.

\vspace{-7pt}
\subsection{Qualitative Analysis}
% We conducted a qualitative analysis by 
% We measured the agreement between the designers' intended pattern and the participants' identified pattern for each chart. We calculated agreement as the percentage of participants selecting the intended pattern for each chart, indicating how consistently participants identified the designers' intended pattern. 
We measured the agreement between the intended pattern and participants’ identified patterns for each chart. We calculated agreement as the percentage of participants who selected the intended pattern, indicating how consistently participants identified it.

\begin{figure*}[t]
    \centering
    \vspace{-30pt}
    \includegraphics[width=\textwidth,height=0.25\textheight,keepaspectratio]{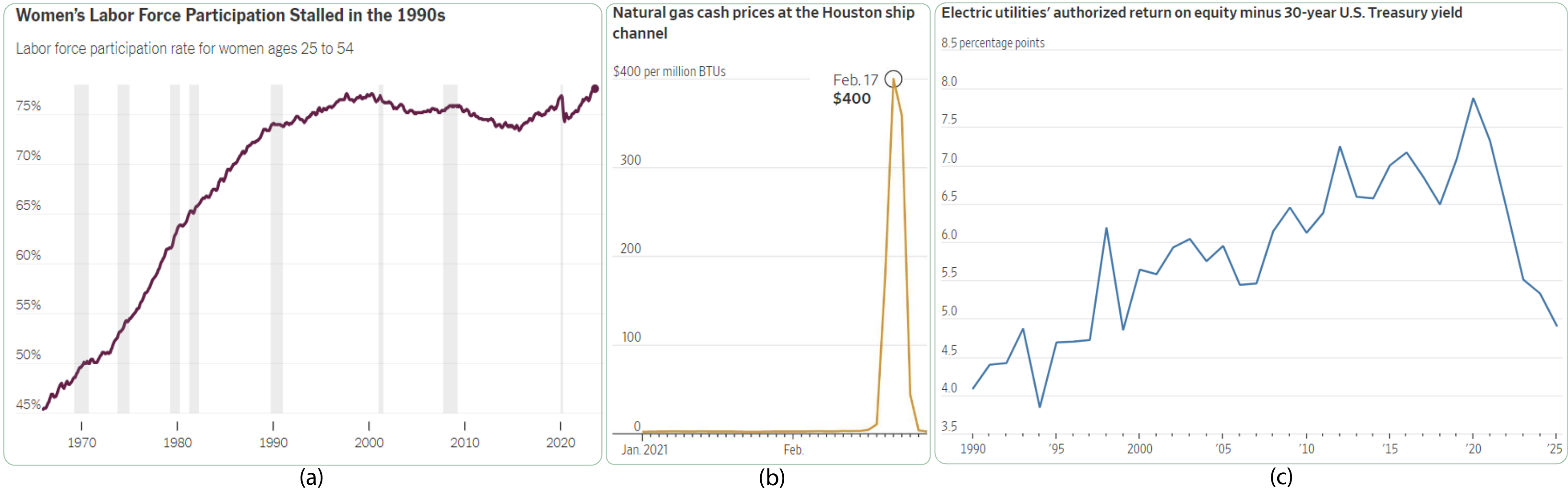}
    \caption{Representative examples discussed in \autoref{sec-discussion}. 
    % (a) A high word count title provides contextual cues that reinforce the upward intended pattern. 
    (a) A chart with a high-word-count title and high agreement with the intended upward pattern.
    (b) A visually salient peak is continuously identified despite a neutral title. (c) An irregular intended pattern with multiple competing visual features elicits diverse pattern interpretations despite strong textual framing.}
    \label{fig:stimuli}
    \vspace{-15pt}
\end{figure*}

% We observed higher agreement among charts with high title word counts. Specifically, 8 of the 9 charts with high word count titles showed above 50\% agreement with the intended pattern, compared to fewer charts containing low or medium word-count titles.
We observed that 9 of the 10 charts with high-word-count titles showed more than 50\% agreement with the intended pattern, compared with fewer charts with low- or medium-word-count titles. Although titles differed in their intended message, the qualitative analysis did not reveal consistent differences in agreement across the neutral, statistical, and affective categories.
% These observations suggest 
% that 
% richer textual framing provides additional contextual cues 
% that complement the visual structure during pattern identification, as illustrated by the high word count title in \autoref{fig:stimuli}a, which achieved high agreement with the intended upward pattern.
These observations suggest an association between high-word-count titles and
greater agreement with the intended pattern; however, title word count alone
does not indicate whether this association reflects additional contextual
information, explicit interpretive cues, or other characteristics of the
stimuli. \autoref{fig:stimuli}a illustrates a high-word-count title that achieved high agreement with the intended upward pattern.

We identified 11 of the 50 charts as containing highly salient visual structures, including prominent peaks, valleys, and strong monotonic trends. Notably, 7 of these 11 charts with salient visual structures showed greater than 50\% participant agreement despite neutral titles or low word counts, suggesting that visually salient structures can guide pattern identification even when textual framing offers limited interpretive cues, as illustrated by the salient peak pattern in \autoref{fig:stimuli}b, which participants consistently identified despite its neutral title.
Beyond overall agreement rates, participant responses revealed recurring interpretive behaviors not evident from statistical analysis alone. Charts with irregular intended patterns frequently elicited multiple plausible interpretations, with participants distributing their responses across multiple (upward, peak, valley, and downward) categories rather than converging on the intended pattern. In contrast, charts containing a single visually dominant structure generally produced more concentrated responses, even when accompanied by neutral or descriptive titles. 
These observations indicate that 
% visually ambiguous patterns 
charts with competing visual features
elicit more diverse interpretations than charts containing a single dominant visual structure.

Agreement also varied across charts with similar title framing characteristics. 
While charts with high title word counts generally corresponded with higher agreement, some still elicited multiple plausible pattern interpretations. These cases
% demonstrate that richer textual framing alone does not guarantee 
show that high-word-count titles do not necessarily correspond to
consistent pattern identification when the underlying visual structure affords several reasonable interpretations, as illustrated by the irregular chart in \autoref{fig:stimuli}c, where participants distributed their responses across multiple pattern categories despite a statistical, high-word-count title. Conversely, even neutral titles produced high agreement when the chart contained a single visually salient pattern.
% Collectively, these observations suggest that quick-view pattern identification arises from the interaction between title framing and visual structure rather than either factor alone.
Collectively, these observations suggest that both title framing and visual structure are associated with quick-view pattern identification, although their individual contributions cannot be isolated in the current stimulus set.
 \vspace{-2pt}

\label{sec-results}

\vspace{-10pt}

\section{Discussion}

Our findings suggest that both title framing and visual structure contribute to pattern identification during quick-view chart interpretation, supporting previous work.~The quantitative analysis showed that title word count and intended message significantly influenced pattern identification, while the qualitative analysis revealed how these textual characteristics interact with salient visual structure to shape viewers' interpretations.
\newline

\vspace{-5pt}
\noindent \textbf{Title Framing Affects Pattern Takeaway:} 
Title framing influences the pattern takeaways that readers form during quick-view chart interpretation.
Our findings demonstrate that charts with high-word-count titles generally achieved higher agreement with the intended pattern, indicating that textual framing provides contextual cues that can reinforce viewers' interpretations. For example, in \autoref{fig:stimuli}a, the chart with a high word count title achieved 83.3\% agreement with the intended upward pattern.
Titles, therefore, contribute to pattern identification by providing contextual information that shapes how visual patterns are identified, consistent with prior work on textual framing in visualization \cite{stokes2022striking}.
Our results suggest that titles serve as more than descriptive labels, providing interpretive cues that guide viewers' understanding of visual patterns. 
As a result, the design and phrasing of chart titles play an important role in visualization communication by guiding viewers' interpretations and influencing the takeaways from a chart.

\noindent \textbf{Salient Visual Structure Guides Pattern Identification:}
% Our findings suggest that 
Salient visual structures play an important role in pattern identification and helping readers interpret charts. 
This finding is consistent with prior work showing that viewers prioritize visually salient structures such as trends, peaks, and valleys when interpreting line charts \cite{proma2025evaluatinglinechartstrategies,proma2025stenography}.
Many of the line graphs in our stimulus pool contained visually prominent patterns, and readers' identified patterns were strongly influenced by these structural characteristics.
For example, the chart in \autoref{fig:stimuli}b contains a prominent peak accompanied by a neutral title that provides little interpretive guidance. Despite the limited textual framing, participants consistently identified the intended peak pattern, suggesting that
% certain 
visual structures can be sufficiently salient to guide pattern identification even without strong textual cues.
%
% \newline
% \vspace{-5pt}

\noindent\textbf{Appropriate Title Framing Complements the Visual Structure:}
While 
% richer textual framing with high word count 
high word count titles generally supported higher agreement; we also observed variability across charts with similar title framing characteristics. \autoref{fig:stimuli}c illustrates a chart with a statistical, high word count title but relatively low agreement with the intended irregular pattern. Although the title provides contextual information about the data being visualized, it offers limited guidance regarding the intended visual pattern.
Participants instead distributed their responses across several plausible patterns, including upward, peak, valley, and downward, reflecting the chart's multiple competing visual features~\cite{proma2025stenography}.~Our findings support prior work showing that textual framing complements visual structure in shaping chart interpretations \cite{kim2021chartsCaptions, stokes2022striking}.~These observations suggest
that when the underlying visual structure affords multiple interpretations, textual framing, specifically the intended message, should be tailored to provide more cues to support pattern identification, and future research should focus on the identification of optimal message framing.
Therefore, quick-view pattern identification arises from the interaction between title framing and visual structure, where textual cues reinforce
the information conveyed in the chart itself.
% \cite{kim2021chartsCaptions, stokes2022striking}
\newline

 \vspace{-8pt}

\noindent\textbf{Limitations and Future Work:}
Our work is limited to chart titles, whereas real-world visualizations often include captions and annotations that also contribute to textual framing. Subtitles treated in our work as titles require separate analysis. Real-world titles are also frequently informative, explicitly naming the depicted pattern, yet such titles were rare in our stimulus set (2 of 50). Our design further cannot isolate each framing category's contribution. Future work should independently manipulate individual title characteristics using a stimulus set balanced across these properties and extend the investigation to quick-view pattern identification in more diverse and complex visualizations.

\label{sec-discussion}

% \vspace{-7.5pt}
\section{Conclusion}
% \fix{add a concise conclusion}
% \noindent\textbf{Future Work:}
% this is from eurovis
% Our study demonstrates that chart title framing influences pattern identification in single-class line charts, whereas highly salient visual patterns remain robust to framing, suggesting that pattern identification may depend on additional factors beyond textual framing. We view this as an initial step; future work is needed to incorporate additional factors and greater diversity in stimulus selection.

% Our study demonstrates that title framing influences pattern identification in single-class line charts, while salient visual structures can guide consistent interpretation even when textual framing provides limited cues. These findings suggest that quick-view pattern identification depends on the interaction between title framing and visual structure. We view this work as an initial step toward broader investigations of textual framing in visualization, incorporating additional textual factors and more diverse chart stimuli.

Our study indicates that title framing 
may
influence pattern identification in single-class line charts, while salient visual structures may guide pattern identification even when textual framing provides limited cues, suggesting that quick-view pattern identification depends on the interaction between title framing and visual
structure. As an initial step toward a broader understanding of textual framing in visualization, this work motivates comprehensive future investigations incorporating additional textual elements, more diverse chart types, and richer visualization contexts.
% \label{sec-discussion}

%\bibliographystyle{abbrv}
% \bibliographystyle{abbrv-doi}
%\bibliographystyle{abbrv-doi-narrow}
\bibliographystyle{abbrv-doi-hyperref}
\newpage
\bibliography{bibliography}
\end{document}